\documentclass[12pt]{article}

\usepackage{epsfig,amsmath,amssymb,graphics,graphicx} 

\begin{document}
%%%%%%%%%%%%%%%%%%%%%%%%%%%%%%%%%%%%%%%%%%%%%%%%%%%%%%%%%%%%%%

\begin{center}
\vskip 7mm
{\large 
Progress in Fractional Differentiation and Applications. \\
2017. Vol. 3. No. 1. P. 1-7. DOI: 10.18576/pfda/030101 }
\vskip 12mm

{\bf \large Economic Interpretation of Fractional Derivatives} \\ 

\vskip 7mm
{\bf \large Valentina V. Tarasova} \\
\vskip 3mm

{\it Lomonosov Moscow State University Business School, \\ 
Lomonosov Moscow State University \\
Moscow 119991, Russia} \\
{E-mail: v.v.tarasova@mail.ru} \\

\vskip 7mm
{\bf \large Vasily E. Tarasov} \\
\vskip 3mm

{\it Skobeltsyn Institute of Nuclear Physics,\\ 
Lomonosov Moscow State University, \\
Moscow 119991, Russia} \\
{E-mail: tarasov@theory.sinp.msu.ru} \\

\begin{abstract}
An economic interpretation of
the Caputo derivatives of
non-integer orders is proposed.
The suggested economic interpretation of 
the fractional derivatives is based 
on a generalization of average and marginal values 
of economic indicators.
We formulate an economic interpretation by using 
the concept of the T-indicator 
that allows us to describe economic processes with memory.
The standard average and marginal values of indicator are special cases of the proposed T-indicator, when the order is equal to zero and one, respectively. 
The fractional derivatives are interpreted as economic characteristics (indicators) that are intermediate between the standard average and marginal values of indicators.
\end{abstract}

\end{center}

\noindent
\noindent
%%%Math. Subj. Classification 2010: 
%%% Fractional derivatives and integrals \\
%%% Perturbation and fractional calculus methods \\


\newpage

%%%%%%%%%%%%%%%%%%%%%%%%%%%%%%%%%%%%%%%%%%%%%%%%%%%%%%%%%%%%%
\section{Introduction}

The theory of derivatives of non-integer orders 
\cite{SKM,Kiryakova,KST,Podlubny,FC2}
has been formulated by famous mathematicians such as Riemann, Liouville, Sonine, Letnikov, Gr\"unwald, 
Marchaud, Weyl, Riesz, Caputo and others. 
Fractional-order derivatives have a wide application
in physics and mechanics, 
since it allow us to describe systems, media and fields
that are characterized by non-locality and memory 
of power-law type.
Recently the fractional derivatives and integrals
are used to describe financial processes \cite{FE1}-\cite{FE5},  
economic processes with nonlocality \cite{EM1}-\cite{EM3}
and economic processes with memory \cite{TT1}-\cite{TT5}.
There are different interpretations of the fractional derivatives and integrals \cite{Int-1}-\cite{Int-F6} such as 
probabilistic interpretations 
\cite{Int-P1}-\cite{Int-P3},
informatic interpretation \cite{FI2016},
geometric interpretations 
\cite{Int-G1}-\cite{Int-G-T} and \cite{Int-GF1}-\cite{Int-GF4},
physical interpretations 
\cite{Int-GF1}-\cite{Int-GF4} and \cite{Int-F1}-\cite{Int-F6}.
In this paper, we propose an economic interpretation
of the fractional derivatives.

%%%%%%%%%%%%%%%%%%%%%%%%%%%%%%%%%%%%%%%%%%%%%%%%%%%%%%%%%%%%%
\section{Economic meaning of derivatives}

Before proceed to consider an economic meaning of derivatives of noninteger order, we will discuss some aspects of the economic interpretation of the standard derivative of the first order.

The economic meaning of the derivative of the first order is that it describes an intensity of changes of an economic indicator
regarding the investigated factor by assuming that other factors remain unchanged. First-order derivative of the function of a indicator defines the marginal value of this indicator. 
The marginal values shows a growth of the corresponding 
indicator per unit increase of the determining factor. 
This is analogous to the physical meaning of speed 
(The physical meaning of speed is the path length travelled per unit of time).
In economic theory, the main marginal values of indicators 
are marginal product, marginal utility, marginal cost, 
marginal revenue, marginal propensity to save and consume, marginal tax rate, marginal demand and some others.

In the study of economic processes is usually performed by calculation of the marginal and average values of indicators 
that are considered as functions of 
the determining factors. Let us give the standard definition of average and marginal values of indicators. Let $Y = Y (X)$
be a single-valued function, which describes the dependence of an economic indicator $Y$ by a factor $X$. Average value $AY_{X}$
of the indicator $Y$ is defined as the ratio of the indicator function $Y = Y (X)$ 
to the corresponding value of factor $X$ as 
\begin{equation} \label{1}
AY_{X} :=\frac{Y(X)}{X} .
\end{equation}

If we have a single-valued and differentiable function $Y = Y (X)$, which describes the dependence of an economic indicator $Y$ on a factor $Y$, the marginal value $MY_{X}$ of this indicator 
is defined as the first derivative of the function $Y = Y (X)$ by the factor $X$ in the form
\begin{equation} \label{2}
MY_{X} := \frac{dY(X)}{dX} .
\end{equation}
It is well-known that the average and marginal values 
are identical ($MY_{X}=AY_{X}$) if the dependence $Y=Y(X)$ 
is directly proportional ($Y(X)=C \cdot X$).

The most important condition of the applicability of 
equation (\ref{2}), which defines the marginal value of indicator, is the assumption that the indicator $Y$ can be represented as a single-valued function of factor $X$. In general, this assumption is not valid \cite{TT6,TT7}, and the dependence of $Y$ on $X$ is not the unique, that is a single value $X$ may correspond to a number of different values of $Y$.

Let us give an example of multivalued dependency 
of an indicator $Y$ on a factor $X$, using results of our article \cite{TT7}. We can consider the indicator and the factor 
as functions of time $t$ that are given by the equations
\begin{equation} \label{3}
X(t)=0.001 \cdot t^2-0.2 \cdot t+70, 
\end{equation}
\begin{equation} \label{4}
Y(t)=0.01 \cdot t^2 -3 \cdot t+1400.
\end{equation}
The dependence of this indicator $Y$ on the factor $X$ is represented by Fig. 1.

%%% ----------------------- PLOTS -------------------------

%%%\newpage
%%%\setcounter{figure}{1}
%%%%%%%%%%%%%%%%%%%%%%%%%%%%%%%%%%%%%%%%%%%%%%%%%%%%%%%%%%%%%%
\begin{figure}[h]
\begin{minipage}[h]{0.47\linewidth}
\resizebox{16cm}{!}{\includegraphics[angle=-90]{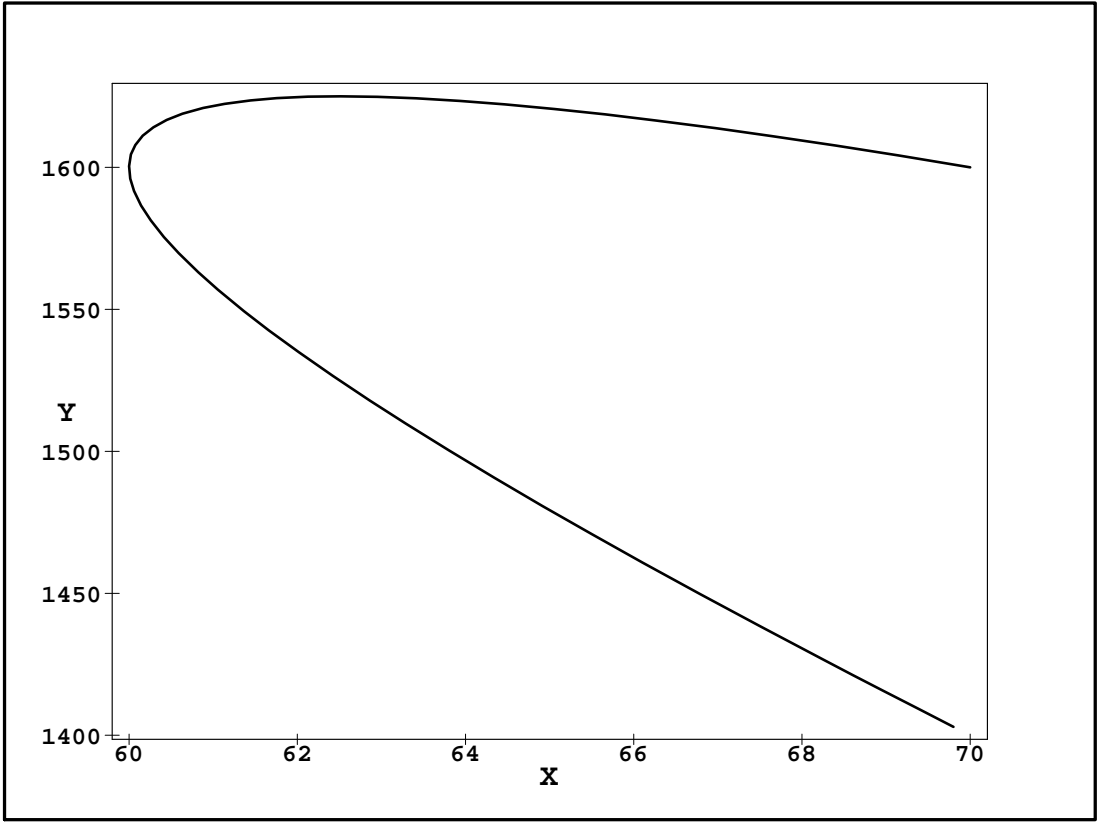}} 
\end{minipage}
\caption{The dependence of $Y$ on $X$, which are
defined by equations (\ref{3}) and (\ref{4}) for $t \in [0,200]$. 
The value $X(t)$ is plotted along the X-axis, 
and the value $Y(t)$ is plotted along the Y-axis.} 
\label{Plot1}
\end{figure}

Let us give a second example of multivalued dependency 
of an indicator $Y$ on a factor $X$, using results of our article \cite{TT7}. We can consider the indicator and 
the factor as functions of time $t$ that are given by the equations
\begin{equation} \label{5}
X(t)=8.2 \cdot 10^{-9} \cdot t^4-
1.5 \cdot 10^{-5} \cdot t^3 + 5.4 \cdot 10^{-3} t^2-
0.58 \cdot t+70, 
\end{equation}
\begin{equation} \label{6}
Y(t)=7.5 \cdot 10^{-6} \cdot t^4-
3.5 \cdot 10^{-3} \cdot t^3+0.51 \cdot t^2-24 \cdot t+1700.
\end{equation}
The dependence of this indicator $Y$ on the factor $X$ 
is represented by Fig. 2.

%%% ----------------------- PLOTS -------------------------

%%%\newpage
%%%\setcounter{figure}{1}
%%%%%%%%%%%%%%%%%%%%%%%%%%%%%%%%%%%%%%%%%%%%%%%%%%%%%%%%%%%%%%
\begin{figure}[h]
\begin{minipage}[h]{0.47\linewidth}
\resizebox{16cm}{!}{\includegraphics[angle=-90]{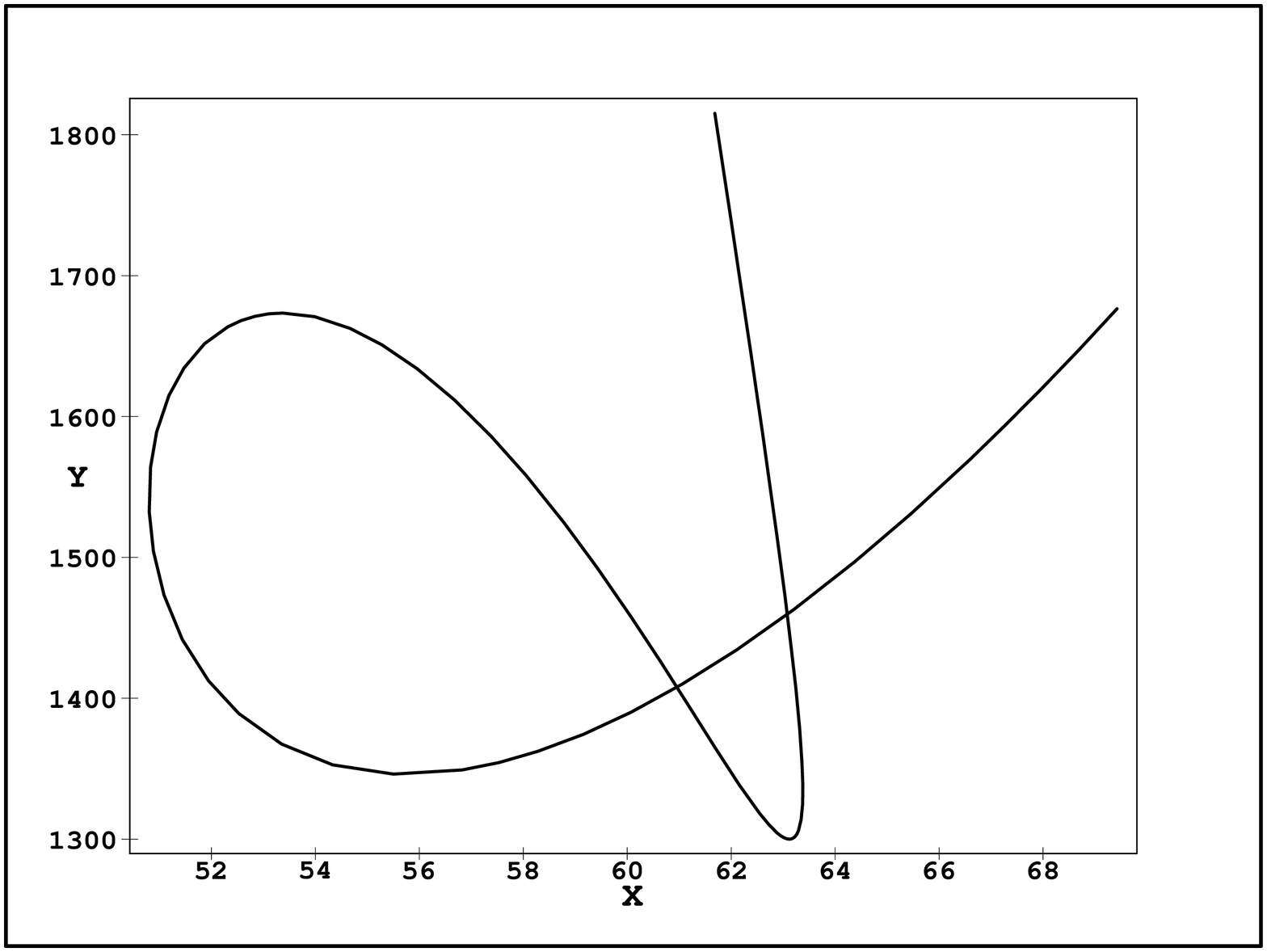}} 
\end{minipage}
\caption{The dependence of $Y$ on $X$, which are
defined by equations (\ref{5}) and (\ref{6}) for $t \in [0,240]$. 
The value $X(t)$ is plotted along the X-axis, 
and the value $Y(t)$ is plotted along the Y-axis.} 
\label{Plot2}
\end{figure}

From graphs presented in Fig. 1 and Fig. 2, it is clear that the value of $X$ may correspond to more than one value of $Y$ in many cases.

If the dependence of the indicator $Y$ on the factor $X$ 
is not single-valued function, we cannot use formulas (\ref{1}) and (\ref{2}) to calculate the average and marginal values of indicator. However, it is possible to avoid this problem of the economic analysis. For this purpose it is sufficient to use the fact that the indicator and factor can practically always be considered as the single-valued functions of time. Therefore to consider the average and marginal values of economic indicators it is necessary to use parametrical dependence of an indicator $Y$ on a factor $X$ in the form of the single-valued functions $Y=Y(t)$ and $X=X(t)$, where the parameter $t$ is time. As a result the average and marginal values of indicators for $t=T$ can be defined by the equations
\begin{equation} \label{7}
AY_{X}(T):=\frac{Y(T)}{X(T)} .
\end{equation}
\begin{equation} \label{8}
MY_{X}(T):= \left(\frac{dY(t)/dt}{dX(t)/dt} \right)_{t=T}= 
\frac{dY(T)/dT}{dX(T)/dT} ,
\end{equation}
where we assume $X(T) \ne 0$ and $dX(T)/dT \ne 0$.

If the dependence of $Y (t)$ on $X (t)$ can be represented as a single-valued differentiable function $Y = Y (X)$, by excluding the time parameter $t$, then expressions (\ref{2}) and (\ref{8}) are equivalent owing to the chain rule. Note that or this purpose it is sufficient that the function $X = X(t)$ be reversible. In this case, equations (\ref{1}) and (\ref{7}) will also be equivalent.

In the differential calculus, the variable $Y$ as function of argument $X$ is called the given parametrically if both of these variables are functions of a third variable t. In our case, the third variable is time $t$. We emphasize that formula (\ref{8}) is a standard definition of the parametric derivative of first order if the function $X = X (t)$ has an inverse in a neighborhood of $t = T$ and the functions $X = X (t)$ and $Y = Y (t)$ have the first derivatives. Moreover, equation (\ref{8}) can be considered as a generalization of the parametric derivative of the indicator $Y = Y (t)$ by the factor $X = X (t)$ at time $t = T$, if $dX (T) / dT\ne 0$. As a result, equations (\ref{7}) and (\ref{8}) can be used for parametric dependences given by equations (\ref{3}), (\ref{4}) and/or (\ref{5}), (\ref{6}). Expressions (\ref{1}) and (\ref{2}) cannot be used for these dependencies. 

As a result, we can state that the economic interpretation of the derivative of the first order for $Y$ on $X$ is associated with marginal indicator (\ref{8}). Equation (\ref{8}) allows us to interpret the first derivative as a growth of indicator $Y$ per unit increase of the factors $X$ at a given time point $t = T$.

It should be noted that one of the main sources and causes
of generation of violation of single-valued property of 
the indicator function $Y(X)$ is the presence of memory in economic processed \cite{TT6,TT7}. Economic agents may remember previous notable changes of the indicator $Y (t)$ and the factor $X(t)$. The memory gives that at repeated similar changes the agents can react on these changes in a different way than it did before. As a result, the indicator will be different for the similar values of the factor. This leads to the fact that the same value of the factor corresponds to a number of different values of the indicator, which is described as a many-valued function $Y = Y (X)$.

Memory of economic agents leads to the fact that the marginal indicator at the time $t = T$ can depend on any changes of $Y (t)$ and $X(t)$ on a finite time interval $(0, T)$, preceding the considered moment $t = T$. Average and marginal values of indicator, which are defined by equations (\ref{7}) and (\ref{8}), depend only on the given time $t = T$ and its infinitesimal neighborhood. Therefore, we can say that the standard definition of these values of indicator, given by formulas (\ref{7}) and (\ref{8}), are applicable only on the condition that all economic agents have full amnesia. Obviously, this approach cannot always be used in the economic analysis. Mathematically, this approach caused by the use of derivatives of integer orders.

In modern mathematics it is known concept of derivative (integrî-differentiation) of non-integer orders \cite{SKM,Kiryakova,KST,Podlubny,FC2}. This concept is used in the natural sciences to describe the processes with memory.
Recently, non-integer order derivatives have been used to describe the financial processes \cite{FE1,FE2,FE3,FE4,FE5}
and the economic processes with memory 
and \cite{TT1,TT2,TT3,TT4,TT5}. There are various types of derivatives of non-integer orders. In this article, we will consider the Caputo fractional derivative. One of the distinguishing features of this derivative is that its action on the constant function gives zero. The use of the Caputo derivative in the economic analysis produces zero value of marginal indicator of non-integer order for the function of the corresponding indicator. There are left-handed and right-handed Caputo derivatives. We will consider only the left-sided derivatives, since the economic process at time $T$ depends only on state changes of this process in the past, that is, for $t <T$, and the right-sided Caputo derivative is determined by integrating the values of $t> T$. Left-sided Caputo derivative of order $\alpha >0$ is defined by the formula
\begin{equation} \label{9}
 _0^CD^{\alpha}_T f(t):= \frac{1}{\Gamma(n-\alpha)} 
\int^T_0 \frac{f^{(n)} (t) \, dt}{(T-t)^{\alpha-n+1}} ,
\end{equation}
where $\Gamma (z)$ is the gamma function, and $T> 0$, 
$n: = [\alpha] +1$, and $f^{(n)}(t)$ is standard derivative 
of integer order $n$ of the function $f (t)$ at time t. For integer values of 
$\alpha = n \in \mathbb{N}$, the Caputo derivative coincides [2, p. 92] with standard derivatives of order $n$, i.e.
\begin{equation} \label{10}
 _0^CD^{n}_T f(t):= \frac{d^nf(T)}{dT^n} , \quad
 _0^CD^{0}_T f(t):= f(T) .
\end{equation}

Using the left-hand Caputo derivative, we can define a generalization of the concepts of marginal and average values of indicator, which allows us to take into account the effects of memory in the economic process. Let the economic indicator 
$Y = Y (t)$ and determining factors $X (t)$ be functions of time 
$t \in [0; T]$. Then economic T-indicator at time $t = T$, which characterizes the economic process with memory, will be defined by the equation
\begin{equation} \label{11}
MY_{X}(\alpha,T):= 
\frac{\, _0^CD^{\alpha}_T Y(t)}{\, _0^CD^{\alpha}_T X(t)} ,
\end{equation}
where $\, _0^CD^{\alpha}_T$ is the left-sided Caputo derivative of order $\alpha \ge 0$, which is defined by expression (\ref{9}). 
The parameter $\alpha \ge 0$ characterizes the degree of attenuation of the memory about the changes of the indicator and factor on the interval $[0, T]$.

It should be noted that the time $t$ can also be considered
as an factor ($X=t$) for economic indicators.
In this case, the  T-indicator (\ref{11}) can be
written in the form
\begin{equation} \label{11b}
MY_{T}(\alpha,T):= \Gamma(2-\alpha) \, T^{\alpha-1}
\, _0^CD^{n}_T Y(t) ,
\end{equation}
where we use $ \, _0^CD^{\alpha}_T t= T^{1-\alpha}/ \Gamma(2-\alpha) $ \cite{KST}.

Average value of indicator (\ref{7}) takes into account only the values of the indicator and factor at times $0$ and $T$. The marginal value of indicator (\ref{8}) takes into account the changes of the indicator and factor in an infinitesimal neighborhood of the time point $t=T$. The proposed economic indicator (\ref{11}) allows us to describe the dependence of economic processes from all state changes on a finite time interval $[0, T]$. This concept takes into account the dependence of the economic indicator at time $T$, not only 
the values of $Y (t)$ and $X (t)$ at this time ($t = T$). Equation (\ref{11}) takes into account all changes of the indicator $Y (t)$ and the factor $X (t)$ on the finite time interval $[0, T ]$.

Let us consider particular cases of the T-indicator (\ref{11})
of order $\alpha$ for the assumption of the existence an single-valued function $Y = Y (X)$. If $\alpha = 0$, the T-indicator (\ref{11}) defines the standard average value of the indicator
\begin{equation} \label{12}
MY_{X}(0,T):= 
\frac{\, _0^CD^{0}_T Y(t)}{\, _0^CD^{0}_T X(t)} =
\frac{Y(T)}{X(T)} = AY_{X}(T) . 
\end{equation}
For $\alpha = 1$, the T-indicator (\ref{11}) defines the standard marginal value of the indicator
\begin{equation} \label{13}
MY_{X}(1,T):= 
\frac{\, _0^CD^{1}_T Y(t)}{\, _0^CD^{1}_T X(t)} =
\frac{dY(T)/dT}{dX(T)/dT} = MY_{X}(T) , 
\end{equation}
which is defined by derivative of the first order for the indicator $Y = Y (X)$ by a factor $X$. From these equations we can see that the average indicator (\ref{7}) and the marginal indicator (\ref{8}) are special cases of the proposed T-indicator (\ref{11}) of the order $\alpha \ge 0$. The indicator (\ref{11}) generalizes the notion of the average and marginal indicators by including these concepts as special cases. In addition, T-indicator (\ref{11}) allows us to consider not only the average and marginal characteristics of economic processes, but also new characteristics that are intermediate between the average and marginal values. The proposed T-indicator includes the whole spectrum of the intermediate values of the indicator from the average and marginal values.

Note that formula (\ref{11}) can be applied to dependencies $Y$ on $X$, which are given by equations (\ref{3}), (\ref{4}) and/or (\ref{5}), (\ref{6}). It is sufficient to apply the formulas, 
which allow us to calculate the Caputo derivatives of order $\alpha>0$ of the power functions, 
\begin{equation}
_0^CD^{\alpha}_T t^{\beta} = \frac{\Gamma(\beta+1)}{\Gamma(\beta+1-\alpha)} \, T^{\beta - \alpha} , \quad 
(\beta >n-1, \alpha >n-1, t>0) ,
\end{equation}
\begin{equation}
_0^CD^{\alpha}_T t^k = 0, \quad (k=0,1, \dots , n-1) ,
\end{equation}
where $n-1 \le \alpha < n$.

%%%%%%%%%%%%%%%%%%%%%%%%%%%%%%%%%%%%%%%%%%%%%%%%%%%%%%%%%%%%%
\section{Conclusion}

As a result, it can be concluded that the economic interpretation of non-integer order derivative $\alpha \ge 0$ directly related with the concept of T-indicator (\ref{11}). The fractional derivatives can be interpreted as a growth of indicator $Y$ per unit increase of the factors $X$ at time $t = T$ in the economic process with memory of power type.

%%%%%%%%%%%%%%%%%%%%%%%%%%%%%%%%%%%%%%%%%%%%%%%%%%%%%%%%%%%%%%
%%%%%%%%%%%%%%%%%%%%%%%%%%%%%%%%%%%%%%%%%%%%%%%%%%%%%%%%%%%%%%
%%%%%%%%%%%%%%%%%%%%%%%%%%%%%%%%%%%%%%%%%%%%%%%%%%%%%%%%%%%%%%

%%%%%%%%%%%%%%%%%%%%%%%%%%%%%%%%%%%%%%%%%%%%%%%%%%%%%%%%%%%%%%
%%%%%%%%%%%%%%%%%%%%%%%%%%%%%%%%%%%%%%%%%%%%%%%%%%%%%%%%%%%%%%
%%%%%%%%%%%%%%%%%%%%%%%%%%%%%%%%%%%%%%%%%%%%%%%%%%%%%%%%%%%%%%

\end{document}